\documentclass[aps,tightenlines,nofootinbib,preprint]{revtex4}
\usepackage{graphicx}
\usepackage{amsmath}
\usepackage{latexsym}
\usepackage{stmaryrd}
\newcommand{\bear}{\begin{eqnarray}}
\newcommand{\eear}{\end{eqnarray}}
\begin{document}

\title{Renormalon Subtraction from the Average Plaquette and  
the \\ Gluon
Condensate}

\author{Taekoon Lee}
\email{tlee@kunsan.ac.kr}
\affiliation{Department of Physics, Kunsan National University, 
Kunsan 573-701, Korea}
\begin{abstract}
A Borel summation scheme of subtracting 
the perturbative contribution from the average
plaquette is proposed using the bilocal
expansion of Borel transform. 
It is shown that the 
remnant of the average plaquette, 
after subtraction
of the perturbative contribution, 
scales as a dim-4 condensate.
A critical review of the existing 
procedure of 
renormalon subtraction is
presented. 
\end{abstract}

\pacs{}


\maketitle
\section{Introduction}
An old problem in lattice gauge theory is extracting
 the gluon condensate from the average plaquette, which in 
pure SU(3) Yang-Mills theory
has the formal expansion \bear
P(\beta)\equiv\langle 1- \frac{1}{3}\text{Tr} \,\text{U}_{\boxempty}
\rangle=\sum_{n=1}\frac{c_n}{\beta^n} + \frac{\pi^2}{36}
Z(\beta)\langle\frac{\alpha_s}{\pi}GG \rangle a^4 +O(a^6) \,,
\label{ope}
\eear
where $\beta$ denotes the lattice coupling and $a$ the lattice 
spacing.
The difficulty of extracting the gluon condensate is that the average 
plaquette is dominated by the perturbative contribution 
and it is necessary to subtract it
to an accuracy better than one part in $10^4$.
The perturbative coefficients $c_n$ were computed to 10-loop order 
using the stochastic perturbation theory \cite{direnzo2},
 but this alone does not achieve 
the required accuracy. Therefore, any attempt to extract 
the gluon condensate 
using the perturbative expansion must involve 
extrapolation of the perturbative coefficients to higher orders and, 
 the perturbative expansion being asymptotic, proper handling of them.
Since the large order behavior of perturbative 
expansion is determined by
the renormalon singularity of the Borel transform, a natural
 extrapolation scheme 
would be based on the renormalon singularity.  
 A program along this line
was implemented  by Burgio et al., 
and the authors obtained a surprising 
result of power correction that scales as a dim-2 
condensate \cite{direnzo}. This is in contradiction with 
the operator product expansion (OPE) ~(\ref{ope}) that demands 
the leading power correction scale as a dim-4 condensate.

The  claim of the dim-2 condensate was since then
 reexamined by several authors. 
In obtaining the perturbative contribution,
 Horsley et al. employed an extrapolation 
 scheme based on the power law and truncation of  the perturbative 
 series at the minimal element \cite{horsley}, 
 and Rakow  used stochastic
 perturbation with boosted coupling to
 accelerate convergence \cite{rakow},
  and Meurice  employed extrapolations based on assumed singularity of 
  the plaquette in the complex $\beta$-plane as well as the renormalon 
  singularity, with truncation at the minimal element \cite{meurice}.
  All these studies did not see any evidence of a dim-2 condensate but
  found the plaquette data was consistent with
  a dim-4 condensate. 

To help settle these  conflicting views on the dim-2 condensate
we present in this paper  a critical review of the renormalon-based 
approach of  \cite{direnzo}, and   reveal a serious flaw 
in the program of renormalon subtraction, and  show that the plaquette
data,  when properly  handled, is consistent with a dim-4 condensate.

Specifically, we shall show that
the continuum scheme employed for renormalon subtraction in 
 \cite{direnzo} is not at all a scheme 
where the perturbative coefficients follow a renormalon pattern, 
 and therefore the 
claimed dim-2 condensate is severely contaminated by perturbative
 contribution and  cannot be interpreted as a power correction. 
We then introduce a renormalon subtraction scheme based on the bilocal
expansion of  Borel transform, and show that the plaquette data can be
fitted  well by the sum of a  dim-4 condensate and the Borel summed 
perturbative contribution.

\section{Renormalon subtraction by matching  large order behaviors}

In this section we give a critical review on the renormalon subtraction 
procedure of \cite{direnzo}.
 The perturbative coefficients $c_n$ 
 of the average plaquette at large orders
  are expected to
 exhibit the large order behavior of the infrared renormalon
 associated with the gluon condensate,
but the computed coefficients using stochastic perturbation
 theory turn out
to grow much more rapidly than a renormalon behavior. 
This implies that the 
coefficients are not yet in the asymptotic regime, which is
expected to set in around 
at order $\bar{n}=\beta z_0$ ($z_0$ given below
in  (\ref{consts})), which gives $\bar{n}\sim 30$ for $\beta\sim 6$,
 far higher than the computed levels.
It therefore appears  all but impossible  to extract
the gluon condensate directly from using 
the stochastic perturbation theory, since the 
perturbative contribution must be subtracted, 
at least, to orders in the 
asymptotic regime.
 
 In Ref.\cite{direnzo} this problem was 
 approached by introducing a continuum
 scheme in which the renormalon contribution is subtracted by 
 matching the
 large order behavior in the continuum scheme to the computed 
 coefficients  in
 the lattice scheme.
 Specifically, in order to relate $c_n$ of the lattice scheme
  with the renormalon behavior the average 
plaquette
is  written, essentially, as
\bear
P(\beta)=P^{\rm ren}(\beta_c)+ \delta P(\beta_c) +P_{\rm NP}(\beta)\,,
\label{decomposition}
\eear
where 
\bear
P^{\rm ren}(\beta_c)= \int_0^{b_{\rm max}} e^{-\beta_c b}
\frac{\cal N}{(1-b/z_0)^{1+\nu}} db \eear
with $\beta_c$ denoting the coupling in the continuum scheme defined by    
\bear
\beta_c=\beta-r_1-\frac{r_2}{\beta}
\label{beta_rel}
\eear    
and
\bear
z_0=\frac{16\pi^2}{33}\,,\quad \nu=\frac{204}{121}\,.
\label{consts}
\eear
In Eq. (\ref{decomposition}) the plaquette is divided into 
perturbative
contributions, comprised of the renormalon contribution  $P^{\rm 
ren}$ and the rest of the perturbative contribution  $\delta P$,   and 
nonperturbative power correction $P_{\rm NP}$. 
In this splitting, the asymptotically divergent behavior
of the perturbative contribution is contained in $P^{\rm 
ren}$, and  $\delta P$ denotes the rest that can be
 expressed as a convergent series. 
 (Here, the renormalons other than that associated
with the gluon condensate and the subleading singularities at $b=z_0$ 
are ignored, which, if necessary, 
can be incorporated in  $P^{\rm ren}$.)

We now  define $ P_{\rm NP}^{(N)}$ with
\bear
P_{\rm NP}^{(N)}(\beta)\equiv P(\beta)-
P^{\rm ren}(\beta_c)-\sum_{n=1}^{N}
(c_n-C_n^{\rm ren})\beta^{-n}
\label{powercorrection}
\eear
where $C_n^{\rm ren}$ denotes the perturbative coefficients of
$P^{\rm ren}$   in  power expansion in $1/\beta$. Note that
$P_{\rm NP}^{(N)}$ is free of perturbative coefficients to order $N$.
The constants $r_1,r_2$ that define the continuum scheme and the
normalization constant $\cal N$  are determined so that    $C_n^{\rm
ren}$ converges to $c_n$ at large orders. In the continuum scheme with
\bear
r_1=3.1\,, \quad r_2=2.0
\label{scheme}
\eear
and an appropriate value for $\cal N$ it was observed that 
 $C_n^{\rm ren}$ converge to $c_n$ at the orders computed in 
stochastic
perturbation theory. The last term in 
(\ref{powercorrection})  being a 
converging series  $ P_{\rm NP}^{(N)}$
will be well-defined at $N\to\infty$, and  this is precisely the  
quantity that was assumed
to represent the power correction, and it was $P_{\rm NP}^{(8)}$ that 
was shown to scale as a dim-2 condensate.

The essence of this procedure is that the 
isolation of the renormalon contribution is obtained  
by matching the large order 
behaviors in the
 lattice and continuum schemes, in which the matching 
does not involve the low order
 coefficients. Although  the
renormalon-caused large order behaviors of any two schemes can be
matched, independently of the low order coefficients, it must
be noted that the matching  would  work only when the known
coefficients in both schemes  exhibit  renormalon behavior. 
Since, however, the computed coefficients in the lattice scheme are 
far from being in the asymptotic regime  
and do not follow the renormalon pattern
the matching cannot be performed reliably; Therefore, the conclusion 
of a dimension-2 
condensate based on it should be reexamined.

That the above matching has a serious flaw can be  easily shown
by mapping the perturbative coefficients in the  lattice scheme to the
continuum  scheme (\ref{scheme}).
If the latter is 
indeed a good scheme for renormalon subtraction the mapped coefficients
should exhibit a renormalon behavior.
However, as can be seen in Table ~\ref{Table1}, 
which is obtained by mapping
the central values of $c_n$ from the stochastic perturbation theory,
 the coefficients are alternating in sign and 
far from being of a renormalon behavior. This shows 
that when mapping the
perturbative coefficients between the 
lattice scheme and (\ref{scheme}) the relatively high
order coefficients (say, 7-10 loop orders) are still very sensitive
on the low order coefficients.
Therefore, the above large order matching cannot be performed reliably
with the computed coefficients, and   (\ref{scheme})
cannot be the right scheme where one can isolate and subtract
 the renormalon contribution.
 
\begin{table}[t]
\begin{tabular}{cccccccc}\hline\hline
$c_1^{\text{cont}}$&$c_2^{\text{cont}}$&$c_3^{\text{cont}}$&
$c_4^{\text{cont}}$&$c_5^{\text{cont}}$&$c_6^{\text{cont}}$&
$c_7^{\text{cont}}$&$c_8^{\text{cont}}$  \\ \hline
2.0&-4.9792&10.613&-10.200&-44.218&316.34&-1096.&1947.\\ \hline\hline
\end{tabular}
\caption{The perturbative coefficients 
of the average plaquette in the
continuum scheme.}
\label{Table1}
\end{table}

Checking the internal consistency of the 
subtraction scheme based on the 
matching of large order behavior
also shows the underlying problem. 
The nonperturbative term in (\ref{decomposition}) can be written 
using  (\ref{powercorrection}) as
\bear
P_{\rm NP}(\beta)=    P_{\rm NP}^{(N)}(\beta) -\{\delta 
P(\beta_c)-\sum_{n=1}^N(c_n-C_n^{\rm
ren}) \beta^{-n}\}\,.
\eear 
For $  P_{\rm NP}^{(N)}$ to represent the power correction 
it is clear that
\bear
\left|\delta P(\beta_c)-\sum_{n=1}^N(c_n-C_n^{\rm ren}) 
\beta^{-n}\right| \ll
  P_{\rm NP}^{(N)} (\beta)
\label{criterion}
\eear
must be satisfied. Since $\delta P(\beta_c)$ is by definition 
a convergent
quantity it can be written in a series expansion
 \bear
\delta P(\beta_c)\equiv \sum_{n=1}^\infty D_n \beta_c^{-n}\,,
\eear
where $D_n$ can be computed up to the order $c_n$ are 
known, and
 (\ref{criterion}) can be written approximately   as 
\bear
\frac{|\sum_{n=1}^N D_n \beta_c^{-n}-\sum_{n=1}^N(c_n-C_n^{\rm ren})
\beta^{-n}|}{ P_{\rm NP}^{(N)}(\beta)}\ll1 \,.
\label{criterion1}
\eear
Now, in the scheme  of (\ref{scheme}), and  at $N=8$ and 
$\beta=6.0, 6.2$ and  6.4, for example,  the ratios are $69, 59$ and 
42, respectively:  a severe violation of the consistency condition. This
again confirms that (\ref{scheme}) cannot be a scheme suited for 
renormalon  subtraction.

\section{Renormalon subtraction by Borel summation}

It is now clear that one cannot subtract the perturbative contribution 
in the plaquette by mapping the
renormalon-based coefficients in a continuum
scheme to the lattice scheme, and then matching
them with the computed high order coefficients.  
On the other hand, the lesson of our review
suggests that
one must map the known coefficients in the lattice scheme to a 
continuum one  and look for a scheme where
the mapped coefficients follow a
renormalon behavior.

Once such a scheme is found one can  perform
Borel summation to subtract 
perturbative contribution to isolate the power correction. 
Borel summation is especially suited for this purpose, since
it allows a precise definition of the power corrections
 in OPE \cite{david1,david2,svz}.
The nature of the renormalon
singularity, hence of the large order 
behavior of perturbation, was obtained through the 
cancellation of the ambiguities
in Borel summation and power corrections \cite{mueller}. An
extensive review of renormalons can be found in \cite{beneke}.

In this paper we shall assume that such a scheme exists and 
perform  Borel summation  using the scheme of  bilocal
expansion of Borel transform \cite{surviving}. To Borel-sum the 
divergent perturbation to a sufficient accuracy for the
extraction of power correction, one must have an 
accurate description of the Borel transform in the domain
that contains the origin as well as the first renormalon singularity in 
Borel plane. Bilocal expansion is a scheme of reconstructing the
Borel transform in this domain, utilizing the known perturbative
coefficients and properties of the first renormalon singularity.
After Borel-summing the perturbative contribution the sum of the
Borel summation and a dim-4 power correction can be fitted
to the plaquette data. A good fit would  suggest then
the power correction be of dim-4 type.

The  Borel summation using the first N-loop perturbations 
of the plaquette in bilocal expansion in a continuum scheme
is given in the form:
\bear
{P}_{\rm BR}^{\rm (N)}(\beta)=\int_0^{\infty} e^{-\beta_c b}
\left[\sum_{n=0}^{N-1} \frac{h_n}{n!} 
b^n+\frac{\cal N}{(1-b/z_0)^{1+\nu}} \right] db\,,
\label{bilocal}
\eear
where the integration over the renormalon singularity is performed 
with principal value
prescription. The essential idea of the
bilocal expansion is to interpolate
the two perturbative expansions 
about the origin and about the renormalon singularity to
rebuild the Borel transform. By incorporating the renormalon singularity
explicitly in the expansion  it can extend the  applicability
of the ordinary weak coupling expansion to 
beyond the renormalon singularity, 
and this scheme was shown to work well
with static inter-quark potential 
or heavy quark pole mass \cite{surviving,heavyquark}.
Here, ${\cal N}$ denotes the normalization constant 
of the large order behavior
and the coefficients $h_n$ are determined so that the Borel transform
in (\ref{bilocal}) reproduce the 
perturbative coefficients in the continuum scheme when
expanded at $b\!=\!0$;
Thus $h_n$ depends on the continuum perturbative coefficients 
as well as ${\cal N}$. By definition, ${P}_{\rm BR}^{(N)}(\beta)$, when
expanded in $1/\beta$, reproduces the 
perturbative coefficients of the average 
plaquette to N-loop order that were 
employed in building the Borel transform.
For details of the bilocal expansion of Borel transform
we refer the reader to 
\cite{surviving,heavyquark}. 

The power correction can  then be defined by
\bear
{ P}_{\rm NP}^{\rm (N)}(\beta)\equiv {P}(\beta) -
{P}_{\rm BR}^{\rm (N)}(\beta)\,,
\label{pNP}
\eear
which, by definition, has vanishing perturbative expansions to order $N$.

Using the perturbation to 10-loop order of the plaquette 
we compute ${P}_{\rm BR}^{(10)}(\beta)$ in the continuum scheme 
parameterized by 
Eq. (\ref{beta_rel}). Although $\cal{N}$ can be 
computed perturbatively, using the
perturbations of the average plaquette, it is
 still difficult to obtain a reliable
result using the known coefficients, so here 
it will be treated as a fitting parameter.
Thus in our scheme, as in   \cite{direnzo},  
the fitting parameters are $\cal{N}$ and $r_1, r_2$ of
Eq. (\ref{beta_rel}).

\begin{figure}
\includegraphics[angle=0,width=8cm ]{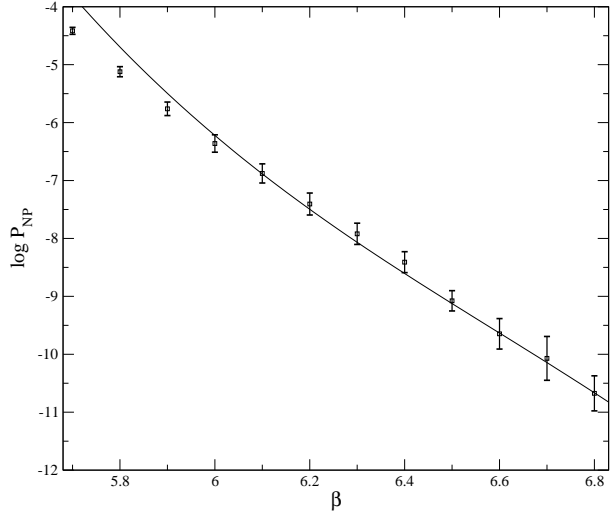}
\caption{ $\log { P}_{\rm NP}$ vs. $\beta$. 
The solid line is for $4 \log(a/r_0) +0.5$. 
The plot shows the power 
correction should be of dim-4 type.}
\label{fig1}
\end{figure}

 Using the plaquette data for 
$6.0\leq \beta \leq 6.8$ from  \cite{plaquette} and
the relation between the lattice spacing $a$ 
and $\beta$ from static quark force 
simulation \cite{sommer}
\bear
\log(a/r_0)=-1.6804 - 1.7331(\beta - 6) + 0.7849(\beta - 6)^2 - 
 0.4428(\beta - 6)^3
 \eear
 the fit gives ${\cal N}=165$ and
 \bear
r_1=1.611, \quad r_2=0.246\,,
\label{fitted}
\eear
which values are substantially 
different from those in (\ref{scheme}).
 The result of the fit is shown in 
 Fig. \ref{fig1}, which shows that the power correction is 
 consistent with 
 a dim-4 condensate. The agreement improves as $\beta$
  increases, albeit with larger uncertainties;
   The  deviation at low $\beta$ ($\beta <6$)
   may be attributed to a dim-6
 condensate, which may be seen, though not presented here, 
  by that adding a 
 dim-6 power correction in the fit  
 improves the agreement in the whole
 range of the plot.   The error bars are from the 
 uncertainty in the simulated
 perturbative coefficients of the plaquette. 
 The uncertainty in the normalization constant
 does not appear to be large: for example, 
 a variation of 20\% in ${\cal N}$ causes 
 less than a quarter of those by the perturbative coefficients.

From the fit we obtain a dim-4 power correction of
 $P_{\rm NP}\approx 1.6\,\, (a/r_0)^4$. 
 Because of the asymptotic nature of the perturbative series 
 the power correction of the plaquette 
 is dependent on the subtraction scheme
 of the perturbative contribution, and 
 thus our result may not be 
 directly compared to those from other 
 subtraction schemes. Nevertheless, it is 
 still interesting to observe that 
the result is roughly consistent 
with $0.4\,\, (a/r_0)^4$ of  \cite{rakow} 
and $0.7 \,\,(a/r_0)^4$ of  \cite{meurice}. 
Our result turns out to be
a little  larger
than those estimates; This may be partly accounted for
by the fact that the existing 
results were from the fit in the low $\beta$ range
of $\beta \lesssim 6$, in which 
range the data are below our fitted curve.
 


\section{Summary}         

We have reexamined the claim of dim-2 
condensate in the average plaquette,
and shown that the renormalon subtraction 
procedure of \cite{direnzo}
 that gave rise to the dim-2 condensate 
 fails consistency checks and 
 cannot be reliably 
 implemented with the known results of 
 stochastic perturbation theory.
We then introduced a renormalon subtraction scheme based on the 
bilocal expansion of Borel transform 
and found that the plaquette data is
consistent with a dim-4 condensate.

\begin{acknowledgments}

This work was  supported in part by 
Korea Research Foundation Grant (KRF-2008-313-C00168). 
\end{acknowledgments}

\bibliographystyle{apsrev}
\bibliography{condensate}

\end{document}